\documentclass[twocolumn,preprintnumbers,amsmath,amssymb]{revtex4}
\usepackage{graphicx}% Include figure files
\usepackage{dcolumn}% Align table columns on decimal point
\usepackage{bm}% bold math

\begin{document}

\title{Comment on: "Current-voltage characteristics and zero-resistance state in 2DEG"}

\author{M. V. Cheremisin}
\affiliation{A.F.Ioffe Physical-Technical Institute,
St.Petersburg, Russia}

%\date{\today}

\begin{abstract}
We demonstrate that N(S)-shape current-voltage characteristics proposed to explain zero-resistance state in Corbino(Hall bar) geometry 2DEG (cond-mat//0302063, 0303530) cannot account essential features of radiation-induced magnetoresistance oscillations experiments.
\end{abstract}

\maketitle

\begin{figure}[tbp]
\begin{center}\leavevmode
\includegraphics[width=0.8\linewidth]{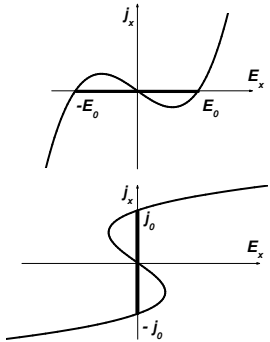}
\caption{\label{fig:CVC} CVC in Corbino(top) and Hall bar(bottom)
geometry suggested in Ref.\cite{Aleiner,Volkov}. The bold lines
depict ZCS and ZRS.}
\end{center}
\end{figure}

Recently, a great deal of interest has been focussed on the
anomalous transport behavior of 2DEG subjected to microwave and
weak magnetic fields \cite {Zudov, Mani}. Possible scenarios for
this phenomenon were discussed in
\cite{Phillips}-\cite{Mikhailov}. As it was shown in
\cite{Durst,Aleiner,Shi}, at certain magnetic field and
irradiation strength the longitudinal conductivity may become
negative argued to result in zero resistance(conductance) states
ZRS(ZCS)observed in experiments. In Ref.\cite{Volkov}
current-voltage characteristic (CVC) of irradiated 2D system in
classically strong magnetic field suggested to exhibit N(S)-shaped
behavior(Fig.\ref{fig:CVC}) in Corbino(Hall bar) geometry
respectively.

We now discuss the relevance of the ZRS model\cite{Volkov}
regarding to experiments \cite{Mani,Yang}. According to
\cite{Aleiner, Volkov}, in ZRS the Hall bar sample exhibits the
current domain structure(bold line in Fig.\ref{fig:CVC},bottom
panel), when $ \left| j \right|<j_{0}$. Upon current enhancement
$\left| j \right|>j_{0}$ the steady current state eventually sets
on. ZRS disappears. In contrast to the above scenario, the
experiments\cite{Mani} demonstrate insensitive nature of the
radiation-induced resistance oscillations within 30-times change
in current magnitude.

Another problem in question is whether ZRS model presents a
transparent physics with respect to Corbino and Hall bar
simultaneous measurements\cite{Yang}. For Corbino geometry sample
CVC argued\cite{Volkov} to be N-shaped(Fig.\ref{fig:CVC}, top
panel). At fixed voltage between the inner and outer contacts
Corbino sample represents the electric field domain structure(ZCS)
when $\left| E \right|<E_{0}$. Let us assume now ZCS at fixed
applied electric field $E=E_{0}$ and certain value of the magnetic
field. Since $\rho_{xx} \simeq \sigma_{xx}/\sigma_{yx}^2=0$, the
system remains ZRS as well. We argue that the experimental
data\cite{Yang} provide the strong evidence that the
radiation-induced conductivity oscillations in Corbino samples are
identical to those deduced from Hall bar measurements data.
Invoking now the formalism\cite{Volkov}, we conclude that at
certain magnetic field and $E=E_{0}$ the ZCS(Corbino) corresponds
to ZRS(Hall bar) at $j=j_{0}$, where $j_{0}=\sigma_{yx}E_{0}$
according to Ref.\cite{Volkov}. Obviously, in real
experiments\cite{Yang} the above relationship is violated since
the applied constant current(Hall bar setup) and constant electric
field(Corbino setup) are presumably independent.

Finally, it turns out that ZRS model is unable to explain some
essential experimental findings\cite{Mani,Yang}.

This work was supported by RFBR(grant 03-02-17588), and
LSF(HPRI-CT-2001-00114, Weizmann Institute)

\end{document}